%% file: main.tex
\documentclass[runningheads]{llncs}

\usepackage[T1]{fontenc}
\usepackage[american]{babel}
\usepackage{graphicx}
\usepackage{microtype}

\usepackage{float}
\usepackage{caption}
\usepackage[table,xcdraw]{xcolor}
\usepackage{booktabs}
\usepackage{siunitx}
\usepackage{stfloats}
\usepackage{wrapfig}
\usepackage{enumitem}
\usepackage{multirow}
\usepackage{subcaption}
\usepackage{url}
\usepackage[hidelinks]{hyperref}
\usepackage{orcidlink}
\usepackage{longtable}
\usepackage{amssymb}
\usepackage{makecell}
\usepackage{array}
\usepackage{nicefrac}

\begin{document}

\title{An Empirical Analysis of High-Performance Computing Education in Germany}

\titlerunning{An Empirical Analysis of HPC Education in Germany}

\author{
Anna-Lena Roth\inst{1}\orcidlink{0000-0002-6463-3486} \and Jonas Posner\inst{1}\orcidlink{0000-0002-1234-5678}
}

\authorrunning{Roth and Posner}

\institute{
Fulda University of Applied Sciences, Fulda, Germany \\
\email{\{anna-lena.roth, jonas.posner\}@cs.hs-fulda.de}
}

\maketitle

% ---------------------------------------------------------------------
% Abstract
% ---------------------------------------------------------------------

\begin{abstract}
\input{sections/00-abstract}

\keywords{HPC \and Parallel Programming \and Education \and Curricula}
\end{abstract}

% ---------------------------------------------------------------------
% Main Content
% ---------------------------------------------------------------------

\input{sections/01-Intro}

\input{sections/02-Methodology}
\input{sections/03-Results}
\input{sections/04-International}
\input{sections/05-Discussion}
\input{sections/06-Conclusion}

% ---------------------------------------------------------------------
% Bibliography
% ---------------------------------------------------------------------

\bibliographystyle{splncs04}
\bibliography{References_short}

\end{document}

%% file: sections/00-Abstract.tex
The increasing relevance of High-Performance Computing (HPC) necessitates the systematic integration of parallel programming and performance-oriented competencies into computational science curricula.
Effective HPC education requires not only theoretical foundations but also hands-on experience with real cluster infrastructures to develop a practical understanding of scalability, efficiency, and architectural distinctions between shared and distributed memory systems.
Although curricular initiatives advocate such integration, comprehensive cross-institutional evidence on how HPC education is implemented---and how curricular structures relate to locally available infrastructure---remains limited.
To examine this relationship, we conducted a systematic empirical assessment of HPC education across 102 academic institutions in Germany.
Based on an analysis of module handbooks and course catalogs, 178 HPC-related courses were identified and evaluated with respect to competency coverage and curricular placement.
While 67.6\% of institutions offer at least one HPC-related course, these offerings are predominantly elective modules situated at the master's level, with comparatively limited integration in bachelor's programs.
To contextualize these findings, local academic HPC cluster infrastructures were additionally assessed regarding availability, size, and documented accessibility for teaching purposes.
Although 61.8\% of the institutions operate HPC clusters, only 23.0\% explicitly document their availability for educational use, as infrastructures are primarily reserved for research.
Statistical analysis reveals a significant association between restricted teaching access and reduced curricular emphasis on practical competencies, including resource management, cluster usage, debugging in parallel environments, and performance analysis.
Overall, the findings indicate a structural imbalance between theoretical instruction and the development of practical HPC competencies within German higher education.

%% file: sections/01-Intro.tex
\section{Introduction} \label{01-introduction}

High-Performance Computing (HPC) has become increasingly important over the past decades.
This development is driven by rapidly growing and increasingly precise data volumes in science and industry, including climate simulations, financial transaction streams, and large-scale training datasets for artificial intelligence.
At the same time, computational models have become more sophisticated and accurate, for example, in deep learning and computational fluid dynamics.
The combination of large-scale data and complex models results in steadily increasing demands on computational performance and memory capacity.

Modern HPC clusters provide the infrastructure required to distribute computations across thousands of interconnected compute nodes.
The performance of such systems is commonly benchmarked and ranked in the Top500 list of the world's fastest supercomputers~\cite{Top500_2025}.
According to this ranking, several powerful systems have reached the exascale level, exceeding $10^{18}$ floating-point operations per second (e.g.,~El Capitan~\cite{ElCapitan2025}, JUPITER~\cite{Jupiter2025}).
However, effectively utilizing HPC clusters requires specialized expertise in parallel programming, cluster architecture, and performance optimization. 
Consequently, HPC has become an essential component of computer science and computational science education.

Several international initiatives and working groups have proposed guidelines and competency frameworks for structuring HPC curricula~\cite{BouvryBrorssonCanal2025,GI_Curriculum,PrasadCharles2023,RajRomanowski2020}.
Despite differences in scope and target audience, the initiatives converge on a common set of core competency areas: (1) parallel programming and parallel programming models, (2) HPC theory and architectural understanding, (3) parallel algorithms and mathematical foundations, (4) middleware and resource management, and above all, (5) theory and practice of performance analysis as a pervasive methodological foundation.

Although international curriculum frameworks for HPC education exist, systematic analyses of their implementation remain limited. 
Most published work reports experiences from individual institutions, resulting in fragmented rather than consolidated insights (e.g.,~\cite{DoepperGrosseLindner2021,RaoofyElisBode2024}). 
A comprehensive overview of how HPC education is currently structured across German academic institutions is lacking.

This paper addresses this gap by presenting a structured empirical study of HPC education at 102 German \emph{universities} and \emph{universities of applied sciences (UAS)}. 
The study investigates (1) whether and how HPC is integrated into computer science-related degree programs, (2) which topics are covered within HPC courses, and (3) which infrastructural and institutional resources are available for teaching.

We first describe the study design and data collection methodology in Section~\ref{sec:03-data_collection}. 
The descriptive analysis of the results is presented in Section~\ref{sec:03-statistics_results}. 
In Section~\ref{sec:03-international_comparison}, we contextualize the findings within the international landscape and discuss related work. 
We then reflect on the implications for HPC education in Section~\ref{sec:03-discussion_hpc_germany}, before concluding the paper in Section~\ref{sec:03-conclusion}.

%% file: sections/02-Methodology.tex
\section{Data Collection and Methodology}\label{sec:03-data_collection}

To assess the prevalence and relevance of HPC education at German academic institutions and the content covered in HPC-related courses, we analyzed the module handbooks and course catalogs of 102 institutions offering computer science degree programs. 
The study was conducted between November and December 2025 and involved a structured examination of the most recent module handbooks valid for the winter semester 2025/2026. 
Both bachelor's and master's degree programs were included.
The sample of 102~institutions was compiled through web-based research to identify accredited computer science programs, and was subsequently verified manually at both the institutional and program level.
The final dataset comprises 64 universities and 38 UAS.
This distribution reflects the structure of the German higher education landscape, in which universities and UAS represent distinct institutional types with different educational profiles.
The raw data underlying this study, including information on the examined institutions, module handbooks, course catalogs, and HPC cluster infrastructures, are publicly available~\cite{RothData2025_HPCGermany}.

For the purpose of this study, HPC-related courses are defined as courses that explicitly address programming models, architectures, and tools designed for developing scalable applications on HPC clusters. 
This includes foundational concepts of HPC cluster architectures, distributed-memory and hybrid systems, communication models, as well as GPU, many-core, and other accelerator technologies.
The definition further encompasses courses that provide practical experience with established HPC programming paradigms, including \emph{Message Passing Interface (MPI)} for distributed-memory systems, %\emph{Open Multi-Processing (
\emph{OpenMP} for shared-memory systems, and \emph{CUDA} or \emph{Open Computing Language (OpenCL)} for GPU programming. 
Courses covering parallel numerical methods, scalable algorithms, and performance analysis and optimization on HPC systems were also included.
Courses primarily focusing on operating-system-level concurrency or general distributed systems were excluded. 
In particular, traditional courses on computer architecture (e.g., von Neumann architecture, pipelining, branch prediction), basic thread-based parallelization mechanisms (e.g., POSIX/Java threads, synchronization primitives, semaphores), or operating system internals were not considered unless they explicitly addressed HPC-specific scalability, communication, or performance aspects. 
Courses that merely utilize HPC clusters (e.g., in machine learning, big data, or cloud computing contexts) without addressing HPC-related programming models or theoretical foundations were also excluded.

For each identified HPC-related course, the following information was systematically recorded:
\begin{itemize}
    \item Institution name,
    \item Course title and type (lecture, seminar, lab, etc.),
    \item Degree program and level (bachelor's, master's),
    \item Elective or mandatory course, and
    \item Course description and learning objectives.
\end{itemize}

In total, 178 distinct HPC-related courses were identified through the examination of module handbooks and course catalogs.
For these courses, content descriptions and stated learning objectives were extracted and subsequently categorized into recurring HPC topics.

Following topics, derived from the analysis of the course descriptions, served as analytical categories for the subsequent content analysis:
\begin{itemize}
    \item \textbf{MPI}: Use of the Message Passing Interface for distributed-memory parallel programming.
    \item \textbf{OpenMP}: Use of OpenMP for shared-memory parallel programming.
    \item \textbf{GPU}: GPU computing in HPC, including CUDA, OpenCL, or comparable frameworks.
    \item \textbf{Resource Management}: Cluster-based job submission, queue management, and scheduler interaction (e.g., SLURM).
    \item \textbf{HPC Theory}: Fundamental HPC concepts and architectures (HPC clusters, supercomputers, networks, memory hierarchies).
    \item \textbf{HPC Cluster Usage}: Hands-on use of HPC clusters or supercomputers in course activities.
    \item \textbf{Storage Systems}: Parallel storage architectures and file systems (e.g., Lustre~\cite{Lustre}, BeeGFS~\cite{BeeGFS}), including parallel I/O.
    \item \textbf{Linux/Unix Command Line}: Shell-based interaction with HPC systems, including secure remote access (e.g., SSH), scripting, environment configuration, and batch-oriented workflows.
    \item \textbf{Performance Algorithms}: Parallel algorithms and data structures with performance focus.
    \item \textbf{Performance Theory}: Theoretical runtime analysis, speedup, scalability, and performance models (e.g., Amdahl's law).
    \item \textbf{Debugging}: Debugging techniques and tools for parallel applications.
    \item \textbf{Performance Practice}: Practical profiling, tracing, and performance optimization (e.g., with Vampir~\cite{KnuepferBrunst2008}).
\end{itemize}

In a second step, we analyzed the existing HPC cluster infrastructures at the respective academic institutions. 
For this purpose, we systematically reviewed the institutional websites to determine whether the institutions operate their own HPC clusters.
The following information was collected:
\begin{itemize}
    \item Name of available cluster(s),
    \item Operator (e.g., institution computing center, department), %, external institution)
    %\item Number of available HPC clusters,
    \item Hardware information (processors, number of nodes and cores, network),
    %\item Hierarchical classification to cluster tier levels (see Section~\ref{subsec:01-lack-of-resources}),
    \item Application areas (e.g., research, teaching, special research groups),
    \item Access conditions (e.g., for staff, students, external users), and
    \item Affiliation of institutions to HPC consortia.
\end{itemize}

%% file: sections/03-Results.tex
\section{Descriptive Statistics and Results}\label{sec:03-statistics_results}

Table~\ref{table:hpc_overview} summarizes the results of the analysis of HPC-related courses at German universities and UAS. 
It reports the number of courses per institution and indicates which of the previously defined HPC topics are explicitly referenced in the corresponding course descriptions. 
The table provides a high-level overview and does not claim completeness, as actual course content and learning outcomes may differ from the official documentation.
Only explicitly mentioned topics were considered.
The table also indicates whether each institution operates its own HPC cluster.

\input{tables/table_hpc_germany.tex}

Overall, 67.6\% of the examined academic institutions offering computer science degree programs provide at least one HPC-related course.
Specifically, HPC-related courses are offered by 21 of the 38 UAS (55.3\%) and by 48 of the 64 universities (75.0\%).
Among the 178 courses analyzed, 102 (57.3\%) are offered at the master's level, 39 (21.9\%) at the bachelor's level, and 37 (20.8\%) are accessible to students from both levels.
The vast majority of courses are electives.
Only 31 courses are mandatory for specific degree programs or study specializations.
Among institutions that include HPC content in their curricula, the module handbooks contain an average of \num{2.6} HPC-related courses.
UAS that offer HPC provide an average of \num{1.5} courses, while universities offer an average of~\num{3.1}.
Notably, FAU Erlangen-Nuremberg~(9), KIT Karlsruhe~(9), TU~Munich~(8), TU Dresden~(7), TU Darmstadt~(6), and Goethe Uni.\ Frankfurt~(6) offer an exceptionally high number of courses and are all universities.

\subsection{Content Analysis of HPC Courses}\label{sec:03-content_analysis}

\begin{figure}[!b]
    \centering
    \includegraphics[width=\textwidth]{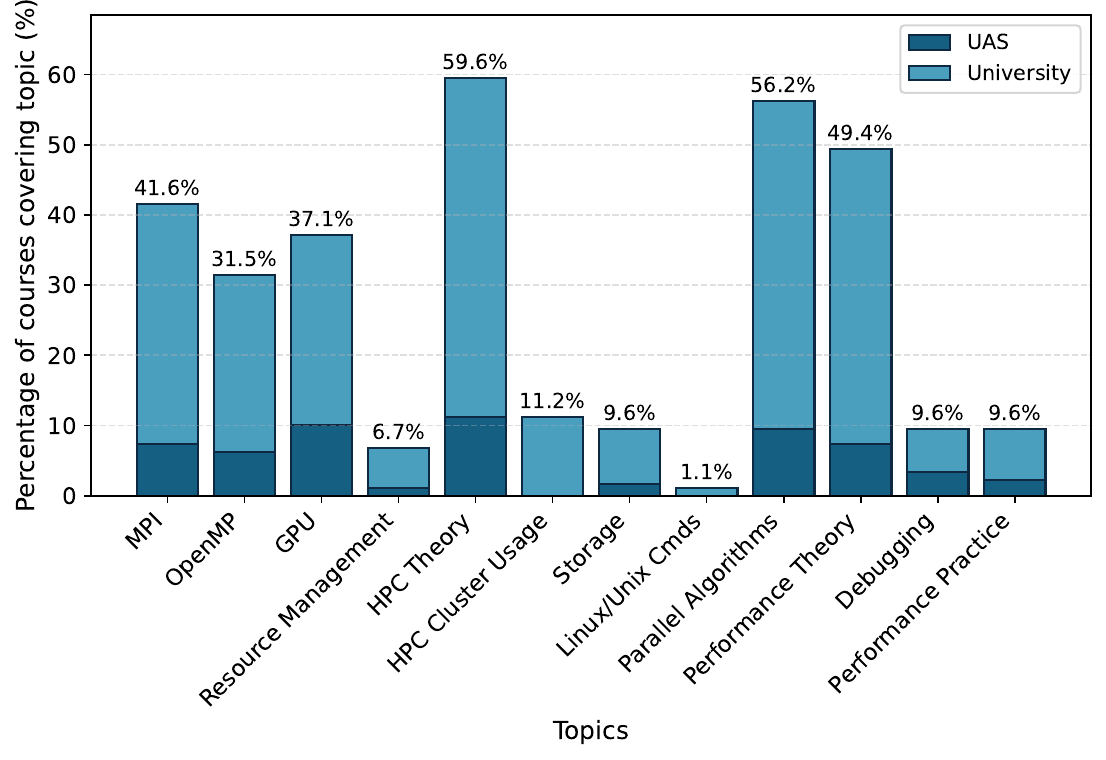}
    \caption{Percentage of HPC courses examined covering HPC topics (n = 178).}
    \label{fig:hpc_topics}
\end{figure}

Figure~\ref{fig:hpc_topics} illustrates the percentage distribution of HPC topics identified through the content analysis of the examined course descriptions.
HPC theory, including the fundamentals of HPC system architectures and functionality, is covered in 59.6\% of the courses.
Topics related to the development of parallel algorithms (e.g., sorting algorithms, numerical methods, and matrix multiplications) appear in 56.2\% of the courses.
Performance-related theory, such as speedup, scalability, and Amdahl's law, is included in 49.4\% of all courses.
Parallel programming models are referenced in 41.6\% of course descriptions for MPI and in 31.5\% for OpenMP, while GPU programming (e.g., CUDA or OpenCL) appears in 37.1\%. 
Some descriptions imply the use of parallel models through references to threads or processes without naming specific frameworks.
Thus, the actual prevalence of MPI, OpenMP, or GPU programming may be higher. 
Explicit HPC cluster usage is documented in only 11.2\% of cases.
Topics addressed in fewer than 10\% of the descriptions include storage systems and parallel I/O~(9.6\%), parallel debugging~(9.6\%), practical performance analysis and optimization~(9.6\%), resource management~(6.7\%), and Linux/Unix command-line usage in HPC contexts~(1.1\%).
No institution covers all defined HPC topics (see Table~\ref{table:hpc_overview}). 
FAU Erlangen-Nuremberg and TU Munich cover the largest number of topics (ten out of twelve each) and also offer the highest number of HPC-related courses.

\subsection{Analysis of HPC Resources}\label{sec:03-hpc_resources}

Of the 102 institutions analyzed, 63 (61.8\%) operate at least one local HPC cluster for research and/or education, based on publicly available information~\cite{RothData2025_HPCGermany}. 
This corresponds to 55 of 64 universities (85.9\%) and eight of 38 UAS (21.1\%). 
In total, 87~HPC clusters were identified (79 at universities and eight at UAS).
Only clusters that could be explicitly assigned to a specific institution were considered.

\begin{figure}[t]
    \centering
    \includegraphics[width=\textwidth]{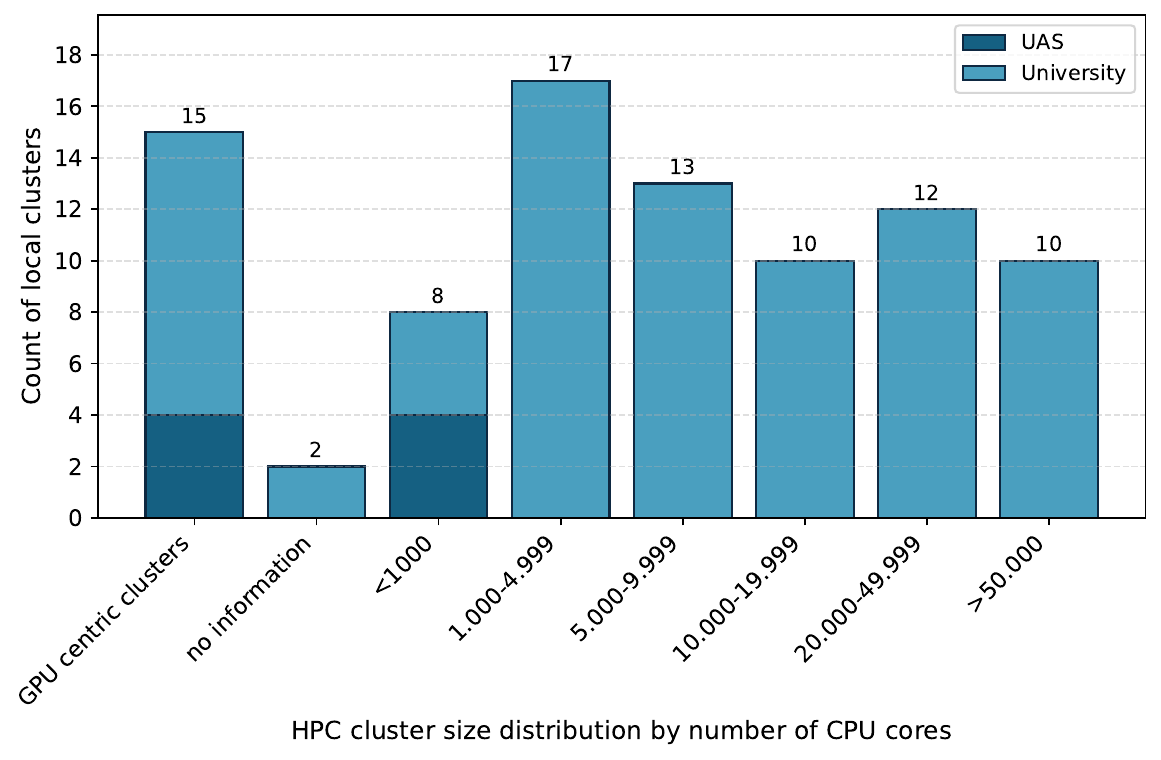}
    \caption{Local HPC clusters at German academic institutions by number of CPU cores (n~=~87).}
    \label{fig:cluster_cores}
\end{figure}

Figure~\ref{fig:cluster_cores} shows the distribution of the identified HPC clusters according to their number of CPU cores.
Clusters are classified by CPU core count.
Exclusively GPU-centric clusters are summarized separately (15 systems).

Eight clusters are classified as small, comprising fewer than \num{1000} CPU cores (four at UAS and four at universities). 
The majority of academic HPC clusters fall within the range of \num{1000} to \num{4999} CPU cores (17 systems, all at universities). 
Beyond this range, cluster sizes increase progressively: 13 universities operate systems with \num{5000} to \num{9999} cores, ten with \num{10000} to \num{19999}, twelve with \num{20000} to \num{49999}, and ten exceeding \num{50000} CPU cores.

Overall, 14 of the 87 identified academic HPC clusters in Germany are listed in the Top500 ranking of the world's fastest supercomputers (as of November 2025)~\cite{Top500_2025}, including five GPU-centric systems. 
Moreover, nine of the ten clusters with more than \num{50000} CPU cores appear in the Top500, indicating a strong correlation between system size and international ranking.
These clusters are Goethe-NHR (Goethe University Frankfurt, rank 181), Otus (Uni.\ Paderborn, rank 191, CPU only), HoreKa (KIT Karlsruhe, rank 202), Emmy (GAU G\"{o}ttingen, rank 250), Noctua~2 (Uni.\ Paderborn, rank 309), Fritz (FAU Erlangen-Nuremberg, rank 350), Lichtenberg (TU Darmstadt, rank 390/409, Phase~1/2), CLAIX-2023 (RWTH Aachen, rank 393, CPU only), and Barnard (TU Dresden, rank 419).

Most clusters are primarily intended for research purposes.
Only 20 of the 87 identified clusters (23.0\%) have explicit educational use documented.
The access conditions for the identified HPC clusters vary significantly:
Ten clusters are exclusively available to specific research groups.
For 49 clusters, an explicit application tied to affiliation with a research project must be submitted.
This could particularly affect students who want to access the clusters for thesis work.
For twelve clusters, outside of internal university exceptions, a proposal that undergoes a peer-review process is necessary to gain access.
Seven clusters are fully or partially accessible to all members of the respective institutions.
For nine clusters, a simple request is sufficient to gain access.

Only a limited number of institutions explicitly document the availability of HPC resources for educational purposes. 
At BTU Cottbus-Senftenberg, a Beowulf PC cluster is designated for educational use, while the main HPC clusters remain reserved for research. 
The University of Bonn provides a small GPU cluster and additionally references a Raspberry Pi cluster for instructional purposes. 
Several universities, including TU Darmstadt, RWTH Aachen, University of Mainz, and TU Dresden, allow limited educational access to partitions of their larger research clusters. 
The University of Paderborn and HS Fulda make decommissioned systems with older hardware available for teaching.

A notable exception is the \emph{bwUniCluster 3.0} operated by KIT Karlsruhe, which is explicitly designed as a shared educational HPC resource for institutions in Baden-W\"urttemberg. 
Through the \emph{bwHPC} network~\cite{bwHPC}, this infrastructure is accessible to multiple universities and UAS, including HS Esslingen, RKU Heidelberg, and University of Ulm.
Despite these initiatives, 14 of the institutions that operate local HPC clusters do not list any corresponding HPC courses in their official module descriptions.

%% file: tables/table_hpc_germany.tex
{\scriptsize
% etwas luftiger: mehr Abstand zwischen Text und Linien
\setlength{\tabcolsep}{4pt}% mehr horizontaler Innenabstand
% besseres Trennen auch in Tabellen
\hyphenpenalty=50      % milde Worttrennung
\exhyphenpenalty=50    % erlaubte Bindestrichtrennung
\pretolerance=10000    % keine zu frühen Overfulls

\begin{longtable}{|>{\raggedright\arraybackslash}p{4cm}|*{14}{c|}}
%\toprule
\hline
\textbf{University} & 
\rotatebox{90}{\parbox{1.85cm}{\centering \textbf{No. courses}}} &
\rotatebox{90}{\parbox{1.85cm}{\centering \textbf{MPI}}} &
\rotatebox{90}{\parbox{1.85cm}{\centering \textbf{OpenMP}}} &
\rotatebox{90}{\parbox{1.85cm}{\centering \textbf{GPUs}}} &
\rotatebox{90}{\parbox{1.85cm}{\centering \textbf{Res. man.}}} &
\rotatebox{90}{\parbox{1.85cm}{\centering \textbf{HPC theo.}}} &
\rotatebox{90}{\parbox{1.85cm}{\centering \textbf{Clus. usage}}} &
\rotatebox{90}{\parbox{1.85cm}{\centering \textbf{Storage}}} &
\rotatebox{90}{\parbox{1.85cm}{\centering \textbf{Linux cmd}}} &
\rotatebox{90}{\parbox{1.85cm}{\centering \textbf{Algorithms}}} &
\rotatebox{90}{\parbox{1.85cm}{\centering \textbf{Perf. theo.}}} &
\rotatebox{90}{\parbox{1.85cm}{\centering \textbf{Debugging}}} &
\rotatebox{90}{\parbox{1.85cm}{\centering \textbf{Perf. prac.}}} &
\rotatebox{90}{\parbox{1.85cm}{\centering \textbf{local clus.}}} 
\\
\hline
\endhead
RWTH Aachen Uni. & 5 & \checkmark &  & \checkmark &  & \checkmark &  & \checkmark &  & \checkmark & \checkmark & \checkmark & \checkmark & \checkmark \\ \hline
HS Aalen & 1 &  &  & \checkmark &  & \checkmark &  &  &  &  & \checkmark &  & \checkmark &  \\ \hline
Uni. Augsburg & 2 &  &  & \checkmark &  & \checkmark &  &  &  & \checkmark &  &  &  & \checkmark \\ \hline
HS Augsburg & 0 &  &  &  &  &  &  &  &  &  &  &  &  &  \\ \hline
OFU Uni. Bamberg & 0 &  &  &  &  &  &  &  &  &  &  &  &  &  \\ \hline
Uni. Bayreuth & 4 & \checkmark & \checkmark & \checkmark & \checkmark & \checkmark &  &  &  & \checkmark & \checkmark &  &  & \checkmark\\ \hline
Freie Uni. Berlin & 1 &  &  &  &  & \checkmark &  &  &  &  &  &  &  & \checkmark \\ \hline
Humboldt-Uni. Berlin & 0 &  &  &  &  &  &  &  &  &  &  &  &  & \checkmark \\ \hline
TU Berlin & 2 & \checkmark & \checkmark &  &  &  & \checkmark &  &  &  &  &  &  & \checkmark \\ \hline
HTW Berlin & 0 &  &  &  &  &  &  &  &  &  &  &  &  & \checkmark \\ \hline
Uni. Bielefeld & 2 &  &  &  &  & \checkmark &  &  &  & \checkmark &  &  &  & \checkmark \\ \hline
RUB Uni. Bochum & 2 & \checkmark & \checkmark &  &  & \checkmark &  &  &  & \checkmark & \checkmark &  &  & \checkmark \\ \hline
HS Bochum & 1 &  &  &  &  &  &  &  &  &  & \checkmark & \checkmark  &  &  \\ \hline
RFWU Uni. Bonn & 4 & \checkmark & \checkmark & \checkmark &  & \checkmark & \checkmark &  &  & \checkmark & \checkmark &  & \checkmark & \checkmark \\ \hline
TU Brunswick & 4 & \checkmark & \checkmark &  &  & \checkmark &  &  & \checkmark & \checkmark & \checkmark &  &  & \checkmark \\ \hline
Uni. Bremen & 0 &  &  &  &  &  &  &  &  &  &  &  &  & \checkmark \\ \hline
HS Bremen & 0 &  &  &  &  &  &  &  &  & 
 &  &  &  &  \\ \hline
HS Bremerhaven & 2 & \checkmark &  &  &  &  &  &  &  & \checkmark &  &  &  &  \\ \hline
TU Chemnitz & 4 & \checkmark &  & \checkmark &  & \checkmark & \checkmark  & \checkmark &  & \checkmark & \checkmark  &  &  &  \\ \hline
TU Clausthal & 1 & \checkmark & \checkmark &  &  & \checkmark &  &  &  & \checkmark &  &  &  & \checkmark \\ \hline
Brand. TU Cottbus-Senftenberg & 0 &  &  &  &  &  &  &  &  &  &  &  &  & \checkmark \\ \hline
TU Darmstadt & 6 &  &  & \checkmark & \checkmark & \checkmark &  & \checkmark  &  & \checkmark & \checkmark & \checkmark & \checkmark & \checkmark \\ \hline
HS Darmstadt & 2 & \checkmark & \checkmark & \checkmark &  & \checkmark &  &  &  & \checkmark & \checkmark & \checkmark & \checkmark &  \\ \hline
TU Dortmund & 2 &  &  & \checkmark &  & \checkmark &  &  &  & \checkmark &  &  &  & \checkmark \\ \hline
FH Dortmund & 1 &  &  &  & \checkmark & \checkmark  &  &  &  &  &  &  &  & \checkmark \\ \hline
HHU Uni. Düsseldorf & 0 &  &  &  &  &  &  &  &  &  &  &  &  & \checkmark \\ \hline
TU Dresden & 7 & \checkmark & \checkmark & \checkmark & \checkmark & \checkmark & \checkmark &  &  & \checkmark & \checkmark &  & \checkmark & \checkmark \\ \hline
HTW Dresden & 1 & \checkmark & \checkmark &  &  & \checkmark &  & \checkmark &  & \checkmark &  & \checkmark &  &  \\ \hline
FAU Uni. Erlangen-Nuremberg & 9 & \checkmark & \checkmark & \checkmark &  & \checkmark & \checkmark & \checkmark &  & \checkmark & \checkmark & \checkmark & \checkmark & \checkmark \\ \hline
FH Erfurt & 0 &  &  &  &  &  &  &  &  &  &  &  &  &  \\ \hline
Uni. Duisburg-Essen & 0 &  &  &  &  &  &  &  &  &  &  &  &  & \checkmark \\ \hline
HS Esslingen & 2 & \checkmark & \checkmark & \checkmark &  &  &  &  &  & \checkmark & \checkmark & \checkmark & \checkmark & \checkmark \\ \hline
Goethe-Uni. Frankfurt & 6 & \checkmark & \checkmark & \checkmark & \checkmark & \checkmark & \checkmark  & \checkmark  &  & \checkmark & \checkmark &  &  & \checkmark \\ \hline
Frankfurt UAS & 0 &  &  &  &  &  &  &  &  &  &  &  &  &  \\ \hline
TU Bergakademie Freiberg & 1 & \checkmark & \checkmark &  &  & \checkmark &  &  &  & \checkmark &  &  &  & \checkmark \\ \hline
ALU Uni. Freiburg & 4 & \checkmark &  & \checkmark &  & \checkmark & \checkmark &  &  & \checkmark &  &  &  & \checkmark \\ \hline
HS Fulda & 1 & \checkmark & \checkmark & \checkmark &  & \checkmark &  &  &  &  &  &  &  & \checkmark \\ \hline
JLU Uni. Gießen & 1 & \checkmark & \checkmark &  &  & \checkmark &  & \checkmark &  &  & \checkmark & \checkmark & \checkmark & \checkmark \\ \hline
TH Mittelhessen & 3 & \checkmark & \checkmark &  &  & \checkmark &  &  &  & \checkmark &  &  &  &  \\ \hline
GAU Uni. Göttingen & 4 & \checkmark & \checkmark & \checkmark &  & \checkmark & \checkmark & \checkmark &  & \checkmark & \checkmark &  & \checkmark & \checkmark \\ \hline
FernUni. Hagen & 3 & \checkmark & \checkmark & \checkmark & \checkmark &  & \checkmark &  &  & \checkmark & \checkmark & \checkmark &  &  \\ \hline
Uni. Hamburg & 3 & \checkmark &  & \checkmark &  & \checkmark &  & \checkmark &  & \checkmark & \checkmark & \checkmark &  & \checkmark \\ \hline
HAW Hamburg & 0 &  &  &  &  &  &  &  &  &  &  &  &  & \checkmark \\ \hline
Leibniz Uni. Hanover & 2 &  &  &  &  & \checkmark &  &  &  & \checkmark &  &  &  & \checkmark \\ \hline
HS Hanover & 1 &  &  &  &  & \checkmark  &  &  &  &  &  &  &  &  \\ \hline
RKU Uni. Heidelberg & 1 & \checkmark &  &  &  & \checkmark &  &  &  & \checkmark & \checkmark &  &  & \checkmark \\ \hline
SUH Uni. Hildesheim & 0 &  &  &  &  &  &  &  &  &  &  &  &  & \checkmark \\ \hline
HS Hof & 0 &  &  &  &  &  &  &  &  &  &  &  &  &  \\ \hline
TU Ilmenau & 0 &  &  &  &  &  &  &  &  &  &  &  &  & \checkmark \\ \hline
TH Ingolstadt & 2 &  & \checkmark & \checkmark & \checkmark & \checkmark &  & \checkmark &  & \checkmark & \checkmark & \checkmark &  &  \\ \hline
FSU Uni. Jena & 5 & \checkmark & \checkmark &  &  & \checkmark &  &  &  & \checkmark & \checkmark &  & \checkmark & \checkmark \\ \hline
RPTU Kaiserslautern-Landau & 1 &  & \checkmark & \checkmark &  & \checkmark &  &  &  & \checkmark & \checkmark &  &  & \checkmark \\ \hline
HS Kaiserslautern & 0 &  &  &  &  &  &  &  &  &  &  &  &  &  \\ \hline
KIT Karlsruhe & 9 & \checkmark & \checkmark & \checkmark & \checkmark & \checkmark & \checkmark & \checkmark &  & \checkmark & \checkmark &  &  & \checkmark \\ \hline
HS Karlsruhe & 0 &  &  &  &  &  &  &  &  &  &  &  &  &  \\ \hline
Uni. Kassel & 3 & \checkmark & \checkmark & \checkmark &  & \checkmark &  &  &  & \checkmark & \checkmark &  &  & \checkmark \\ \hline
HS Kempten & 1 & \checkmark & \checkmark & \checkmark &  & \checkmark  &  &  &  & \checkmark & \checkmark &  &  & \\ \hline
CAU Uni. Kiel & 2 & \checkmark & \checkmark & \checkmark &  & \checkmark & \checkmark &  & \checkmark & \checkmark &  &  &  & \checkmark \\ \hline
FH Kiel & 0 &  &  &  &  &  &  &  &  &  &  &  &  &  \\ \hline
Uni. Koblenz & 1 & \checkmark &  & \checkmark &  & \checkmark &  &  &  & \checkmark & \checkmark &  &  &  \\ \hline
Uni. Cologne & 3 & \checkmark & \checkmark & \checkmark &  & \checkmark &  &  &  & \checkmark & \checkmark &  &  & \checkmark \\ \hline
TH Cologne & 2 & \checkmark & \checkmark & \checkmark &  & \checkmark &  &  &  & \checkmark & \checkmark &  &  & \checkmark \\ \hline
Uni. Konstanz & 0 &  &  &  &  &  &  &  &  &  &  &  &  &  \\ \hline
HS Landshut & 0 &  &  &  &  &  &  &  &  &  &  &  &  &  \\ \hline
Uni. Leipzig & 1 &  &  &  &  & \checkmark &  &  &  & \checkmark &  &  &  & \checkmark \\ \hline
HTWK Leipzig & 0 &  &  &  &  &  &  &  &  &  &  &  &  &  \\ \hline
Uni. Lübeck & 2 & \checkmark & \checkmark & \checkmark &  & \checkmark &  &  &  & \checkmark & \checkmark &  &  & \checkmark \\ \hline
TH Lübeck & 0 &  &  &  &  &  &  &  &  &  &  &  &  &  \\ \hline
OVGU Uni. Magdeburg & 4 & \checkmark & \checkmark & \checkmark &  & \checkmark & \checkmark & \checkmark &  & \checkmark & \checkmark & \checkmark &  & \checkmark \\ \hline
JGU Uni. Mainz & 2 & \checkmark & \checkmark & \checkmark &  & \checkmark &  &  &  & \checkmark & \checkmark &  &  & \checkmark \\ \hline
Uni. Mannheim & 0 &  &  &  &  &  &  &  &  &  &  &  &  &  \\ \hline
TH Mannheim & 3 &  &  & \checkmark &  &  &  &  &  & \checkmark & \checkmark & \checkmark &  &  \\ \hline
PUM Uni. Marburg & 0 &  &  &  &  &  &  &  &  &  &  &  &  & \checkmark \\ \hline
LMU Uni. Munich & 2 & \checkmark & \checkmark & \checkmark &  & \checkmark &  & \checkmark &  & \checkmark &  &  &  & \checkmark \\ \hline
TU Munich & 8 & \checkmark & \checkmark & \checkmark &  & \checkmark & \checkmark & \checkmark &  & \checkmark & \checkmark & \checkmark & \checkmark & \checkmark \\ \hline
HS Munich & 0 &  &  &  &  &  &  &  &  &  &  &  &  &  \\ \hline
Uni. Münster & 3 &  &  & \checkmark &  & \checkmark &  &  &  & \checkmark & \checkmark &  &  & \checkmark \\ \hline
FH Münster & 1 & \checkmark &  & \checkmark &  & \checkmark &  &  &  &  &  &  &  & \\ \hline
TH Nuremberg & 0 &  &  &  &  &  &  &  &  &  &  &  &  & \checkmark \\ \hline
UOL Uni. Oldenburg & 0 &  &  &  &  &  &  &  &  &  &  &  &  & \checkmark \\ \hline
Uni. Osnabrück & 0 &  &  &  &  &  &  &  &  &  &  &  &  & \checkmark \\ \hline
HS Osnabrück & 2 &  &  & \checkmark &  & \checkmark &  &  &  & \checkmark &  &  &  & \checkmark \\ \hline
Uni. Paderborn & 1 &  &  & \checkmark &  & \checkmark & \checkmark  &  &  &  & \checkmark &  &  & \checkmark \\ \hline
Uni. Passau & 1 & \checkmark & \checkmark &  &  &  &  &  &  &  & \checkmark &  &  &  \\ \hline
HS Pforzheim & 1 & \checkmark & \checkmark & \checkmark &  &  &  & \checkmark &  &  & \checkmark &  &  &  \\ \hline
Uni. Potsdam & 4 & \checkmark & \checkmark & \checkmark & \checkmark & \checkmark &  &  &  & \checkmark & \checkmark &  &  & \checkmark \\ \hline
TH Rosenheim & 0 &  &  &  &  &  &  &  &  &  &  &  &  &  \\ \hline
Uni. Rostock & 1 & \checkmark & \checkmark &  &  & \checkmark &  &  &  &  & \checkmark & \checkmark &  & \checkmark \\ \hline
Uni. d. Saarlandes & 0 & &  &  &  &  &  &  &  &  &  &  &  &  \\ \hline
Uni. Siegen & 1 & \checkmark & \checkmark &  &  & \checkmark &  &  &  &  &  &  &  & \checkmark \\ \hline
Uni. Stuttgart & 4 & \checkmark & \checkmark & \checkmark &  &  &  &  &  & \checkmark & \checkmark &  &  & \checkmark \\ \hline
HFT Stuttgart & 1 & \checkmark & \checkmark & \checkmark &  &  &  &  &  & \checkmark & \checkmark & \checkmark & \checkmark & \\ \hline
Uni. Trier & 0 &  &  &  &  &  &  &  &  &  &  &  &  & \checkmark \\ \hline
HS Trier & 1 &  &  & \checkmark &  & \checkmark &  &  &  &  &  &  &  & \\ \hline
EKU Uni. Tübingen & 2 &  &  & \checkmark &  & \checkmark &  &  &  & \checkmark & \checkmark &  & \checkmark & \checkmark \\ \hline
Uni. Ulm & 1 & \checkmark & \checkmark & \checkmark &  & \checkmark &  &  &  &  & \checkmark &  &  & \checkmark \\ \hline
Bauhaus-Uni. Weimar & 1 & \checkmark & \checkmark & \checkmark &  &  &  &  &  & \checkmark &  &  &  & \\ \hline
HS RheinMain & 0 &  &  &  &  &  &  &  &  &  &  &  &  &  \\ \hline
HS Wismar & 0 &  &  &  &  &  &  &  &  &  &  &  &  &  \\ \hline
Bergische Uni. Wuppertal & 5 & \checkmark & \checkmark & \checkmark &  & \checkmark &  &  &  & \checkmark & \checkmark &  &  & \checkmark \\ \hline
JMU Uni. Würzburg & 1 &  &  & \checkmark &  &  &  &  &  &  & \checkmark &  &  & \checkmark \\ \hline
HAW Würzburg-Schweinfurt & 1 & \checkmark & \checkmark & \checkmark &  & \checkmark &  &  &  &  & \checkmark &  &  &  \\ \hline
%\bottomrule
\caption[HPC course offerings by academic institution. Thematic areas are aggregated across all courses offered by each institution]{HPC course offerings by academic institution. Thematic areas are aggregated across all courses offered by each institution. \footnotemark} 
\footnotetext{ 
\scriptsize
\textbf{Abbreviations:}  
Uni.~=~Universität; 
TU~=~Technische Universität; 
HS~=~Hochschule;   
FH~=~Fachhochschule; 
TH~=~Technische Hochschule;    
HFT~=~Hochschule für Technik;
HAW~=~Hochschule für Angewandte Wissenschaften.  
HTW~=~Hochschule für Technik und Wirtschaft;
HTWK~=~Hochschule für Technik, Wirtschaft und Kultur;
UAS~=~University of Applied Sciences.
\\\textbf{International equivalents:}  
Uni.~=~University;  
TU~=~Technical University;  
TH~=~University of Applied Sciences (technical profile);  
HS/FH/HAW/HTW/HFT/HTWK/UAS~=~University of Applied Sciences.
}
\label{table:hpc_overview}
\end{longtable}
}

%% file: sections/04-International.tex
\section{International Comparison}\label{sec:03-international_comparison}

\begin{figure}[!b]
    \centering
    \includegraphics[width=\textwidth]{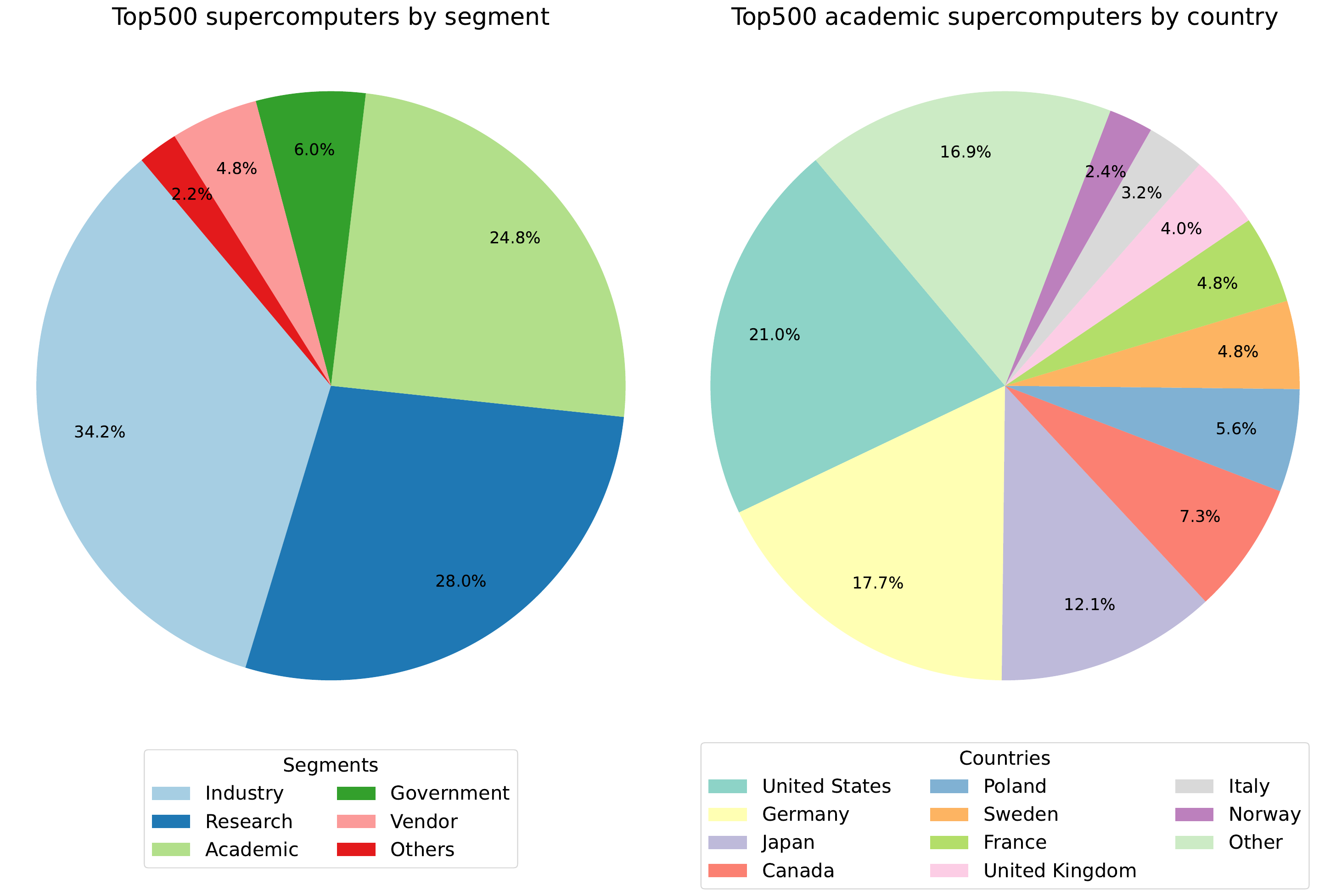}
    \caption{Comparison of HPC segments in the Top500 list (left) and academic supercomputers by country (right). Data source: Top500~\cite{Top500_2025}.}
    \label{fig:hpc_international}
\end{figure}

To contextualize the findings on HPC education at German academic institutions, this section places the results in an international perspective by analyzing data derived from the Top500 list. 
In addition, the findings are compared with international related work on HPC education.

The Top500 list of the world's fastest supercomputers provides not only performance and technical specifications but also information on the institutional segments to which these systems belong. 
Within this classification, the academic segment includes systems directly assigned to universities as well as systems operated by national research centers that provide access to multiple academic institutions.
For example, the SuperMUC-NG~\cite{SuperMuc}, operated at the Leibniz Supercomputing Centre in Munich, is accessible to researchers from various universities through a proposal-based allocation process. 
Furthermore, the Top500 list may represent previously unified systems as separate entries, for instance, when CPU and GPU partitions of a cluster (e.g., CLAIX-2023 at RWTH Aachen) are listed independently.

Overall, 124 of 500 supercomputers (24.8\%) on the Top500 list are assigned to academic institutions worldwide.
Thus, approximately \nicefrac{1}{4} of the world's fastest supercomputers are located at academic institutions or are primarily made available to them.
Figure~\ref{fig:hpc_international} shows the distribution of the academic HPC segments among selected countries.
The largest share of the world's fastest academic supercomputers is held by the USA (21.0\%), Germany (17.7\%), and Japan (12.1\%). 
In total, 25 different countries are represented in the academic category.

In the international context, particularly in the United States, much of the existing literature does not address university-level HPC education but instead focuses on workshops, training programs, and so-called \emph{summer schools} in laboratories~\cite{CallaghanFilingerZhukov2025,DiehlLiJunghans2025,WoffordLueninghoener2020}.
In the U.S., numerous laboratories and research centers operate HPC infrastructures primarily for research purposes and offer associated training initiatives. 
These initiatives typically aim to familiarize researchers with specific systems and usage procedures rather than constituting regular, curriculum-embedded university courses. 
Given their distinct objectives, structures, and target groups, such programs are not considered in the present analysis.

Hebert et al.~\cite{HebertHratischGomesKunkel2024} analyze the availability and use of HPC infrastructure across academic institutions in Wisconsin.
They report a strong concentration of HPC resources at large research universities, while smaller institutions rarely operate independent facilities. 
HPC training is seldom embedded in undergraduate curricula, and active HPC research is largely confined to institutions with local computational resources. 
Comparing their data with 133 international university-based HPC centers, the authors further show that many centers restrict access to internal users and depend heavily on governmental funding. 
They also observe a positive correlation between cluster size and dedicated support staff, underscoring the substantial personnel requirements for operating HPC infrastructures.

Samuel et al.~\cite{SamuelBrennanTonettaSamuelSubediSmith2021} examine the use of the Caliburn HPC cluster at Rutgers University by novice users and derive implications for interdisciplinary HPC education. 
They identify limited resource availability, restricted access, and usability barriers as key structural constraints. 
In their case study, a first-time user spent the majority of the effort on mastering operational prerequisites rather than on the substantive research task, highlighting the considerable overhead associated with HPC usage. 
The authors advocate interdisciplinary instructional models and simplified access mechanisms to reduce these barriers.

Al-Jody et al.~\cite{AljodyTaha2021} survey HPC course offerings at eleven UK institutions. 
Only about half of the analyzed programs include hands-on interaction with HPC clusters, reflecting limited infrastructural capacity. 
To address this challenge, the authors propose a middleware approach based on container technologies to support instructional computing clusters.

Banchelli and Mantovani~\cite{BanchelliMantovani2019} investigate HPC curricula at several European universities. 
While parallel programming models and theoretical performance concepts are consistently covered, practical competencies such as profiling, debugging, energy efficiency, and cluster administration are largely underrepresented. 
They suggest aligning curricula with competencies required for the Student Cluster Competition, including the integration of performance analysis tools such as Extrae~\cite{Extrae} and Paraver~\cite{Paraver}.

%% file: sections/05-Discussion.tex
\section{Discussion}\label{sec:03-discussion_hpc_germany}

Building on the analyses of HPC course content (Section~\ref{sec:03-content_analysis}) and HPC infrastructure availability (Section~\ref{sec:03-hpc_resources}), this section synthesizes both dimensions to examine their structural interdependencies and educational implications.

The results indicate that HPC is broadly embedded in German higher education, yet with pronounced institutional asymmetries.
HPC appears significantly more frequently in university curricula (75.0\%) than at UAS (55.3\%), and universities offer, on average, a substantially higher number of HPC courses (\num{3.1} vs.\ \num{1.5}). 
This disparity corresponds to infrastructural differences: 85.9\% of universities operate local HPC clusters, compared to only 21.1\% of UAS. 
The presence of such infrastructure not only enables practical training but also reflects the availability of specialized personnel capable of supporting HPC education.

HPC-related courses in Germany primarily emphasize theoretical foundations, parallel algorithms, performance theory, and programming with MPI, OpenMP, and CUDA. 
Offerings are predominantly elective modules at the master's level, while bachelor-level integration remains limited.
The five thematic areas proposed by HPC education working groups and initiatives, which are considered essential for comprehensive HPC education (see Section~\ref{01-introduction}), are primarily reflected in the German curricula we examined.
Parallel programming and parallel programming models (1) are well represented, particularly through MPI (41.6\%) and OpenMP (31.5\%), as well as GPU programming with, e.g., CUDA (37.1\%).
HPC theory and architectural understanding (2) are covered in 59.6\% of all courses. 
Parallel algorithms and mathematical foundations (3) are addressed in 56.2\% of all courses. 
Middleware and resource management (4) are included only in 6.7\% for resource management and 9.6\% for storage topics such as I/O and parallel file systems.
In the theory and practice of performance analysis and optimization, 49.4\% of all courses cover performance theory, whereas only 9.6\% include practical performance optimization.

Despite the apparent formal alignment with recommended competency areas, a substantial imbalance between theoretical and practical components emerges.
While performance-theoretical topics such as Amdahl's law and the concepts of speedup, scalability, and load balancing are widely covered, practical skills required to identify and resolve performance problems (e.g., profiling, tracing, optimization) are significantly underrepresented (9.6\%). 
Such competencies are methodologically demanding and typically require access to specialized toolchains and HPC infrastructures, and thus remain difficult to integrate systematically in the absence of appropriate infrastructure.
Explicit HPC cluster usage is documented in only 11.2\% of courses, and practical resource management (6.7\%)---including scheduler-based job submission and allocation via SLURM---as well as debugging in parallel environments (9.6\%) are similarly rare.
These figures indicate that authentic cluster-based workflows are not consistently embedded in curricula.
Linux/Unix command-line proficiency, including shell-based interaction with HPC systems, job scripting, and batch processing, is mentioned in only 1.1\% of the analyzed course descriptions.
Given that many HPC courses are positioned at the master's level, such competencies may be assumed as prior knowledge rather than explicitly taught.
However, the overall low representation of cluster-related topics suggests that numerous courses rely on local programming environments instead of sustained hands-on interaction with HPC systems. 
These findings align with international observations by Banchelli and Mantovani~\cite{BanchelliMantovani2019}, who similarly reported that debugging, complex code optimization, and energy efficiency are underrepresented topics in HPC education.

The infrastructural analysis further reveals a structural accessibility gap.
Although 61.8\% of the analyzed academic institutions operate HPC clusters---some even represented in the Top500 ranking---only 11.2\% of courses explicitly document cluster usage, and merely 23.0\% of clusters are documented as accessible for teaching.
Access is frequently restricted by proposal-based allocation procedures or limited to specific research groups, creating substantial barriers for educational integration.
These patterns are consistent with the findings of Samuel et al.~\cite{SamuelBrennanTonettaSamuelSubediSmith2021}, who identify administrative constraints, access restrictions, and usability hurdles as major impediments to effective HPC adoption in academic contexts.

Taken together, the results suggest that HPC infrastructures in Germany are primarily organized as research resources rather than as systematically integrated educational environments. 
The mere presence of high-performance systems does not automatically translate into accessible, practice-oriented training opportunities. 
This structural configuration particularly disadvantages UAS, where limited local infrastructure constrains the integration of authentic HPC workloads into teaching.

Shared HPC infrastructures, such as the \emph{bwUniCluster 3.0} in Baden-\hspace{0pt}W\"urttemberg, as well as regional and national consortia including HKHLR~\cite{HKHLR}, HPC.NRW~\cite{HPCNRW}, and the nationwide NHR infrastructure~\cite{NHR}, may mitigate access disparities between institutions. 
However, shared systems also introduce governance, scheduling, and administrative constraints that may restrict pedagogical flexibility and institutional autonomy. 
Coordinated resource allocation, limited configuration privileges, and competing research workloads can reduce the feasibility of sustained hands-on instructional use.

Overall, the findings reveal a systemic tension between research-oriented HPC infrastructures and the requirements of competency-oriented HPC education, resulting in structurally limited opportunities for comprehensive practical skill development.

%% file: sections/06-Conclusion.tex
\section{Conclusion}\label{sec:03-conclusion}

This study presents a systematic analysis of HPC education at German academic institutions, integrating curricular and infrastructural perspectives. 
While HPC is widely embedded and key thematic areas are formally represented, a clear imbalance persists between theoretical coverage and practice-oriented competencies.

Although many institutions operate HPC infrastructures, limited educational access and the rare integration of authentic cluster-based workflows restrict opportunities for hands-on experience. 
Practical competencies---particularly Linux command-line skills, resource management, debugging, and performance analysis---remain underrepresented despite their relevance for research and industry.

The findings indicate a structural misalignment between research-oriented HPC infrastructures and competency-oriented educational objectives. 
Strengthening practical HPC education, therefore, requires not only curricular refinement but also sustainable, low-barrier access to HPC systems for teaching.